\documentclass[conference]{IEEEtran}
\IEEEoverridecommandlockouts
\usepackage{cite}
\usepackage{amsmath,amssymb,amsfonts}
\usepackage[normalem]{ulem}

\newcommand*{\affmark}[1][*]{\textsuperscript{#1}}
\usepackage{algorithmic}
\usepackage{graphicx}
\usepackage{textcomp}
\usepackage{xcolor}
\usepackage{booktabs}
\def\BibTeX{{\rm B\kern-.05em{\sc i\kern-.025em b}\kern-.08em
    T\kern-.1667em\lower.7ex\hbox{E}\kern-.125emX}}

\makeatletter
 \let\old@ps@headings\ps@headings
 \let\old@ps@IEEEtitlepagestyle\ps@IEEEtitlepagestyle
 \def\confheader#1{%
 \def\ps@headings{%
 \old@ps@headings%
 \def\@oddhead{\strut\hfill#1\hfill\strut}%
 \def\@evenhead{\strut\hfill#1\hfill\strut}%
 }%
 \def\ps@IEEEtitlepagestyle{%
 \old@ps@IEEEtitlepagestyle%
 \def\@oddhead{\strut\hfill#1\hfill\strut}%
 \def\@evenhead{\strut\hfill#1\hfill\strut}%
 }%
 \ps@headings%
 }
\makeatother

\confheader{%
Accepted by 2023 IEEE International Conference on Communications (ICC), ©2023 IEEE
}

\begin{document}

\title{Beam Selection for Energy-Efficient mmWave Network Using Advantage Actor Critic Learning \\


}

\author{\IEEEauthorblockN{Ycaro~Dantas\affmark[1], Pedro~Enrique~Iturria-Rivera\affmark[1], Hao~Zhou\affmark[1],  Majid Bavand\affmark[2], Medhat Elsayed\affmark[2], \\Raimundas Gaigalas\affmark[2], and  Melike Erol-Kantarci\affmark[1], \IEEEmembership{Senior Member,~IEEE}}
\IEEEauthorblockA{\affmark[1]\textit{School of Electrical Engineering and Computer Science, University of Ottawa, Ottawa, Canada}}  \affmark[2]\textit{Ericsson Inc., Ottawa, Canada}\\
Emails:\{ydant103, pitur008, hzhou098, melike.erolkantarci\}@uottawa.ca, \\\{majid.bavand, medhat.elsayed, raimundas.gaigalas\}@ericsson.com \vspace{-1em}}

\maketitle


\begin{abstract}
The growing adoption of mmWave frequency bands to realize the full potential of 5G, turns beamforming into a key enabler for current and next-generation wireless technologies. Many mmWave networks rely on beam selection with Grid-of-Beams (GoB) approach to handle user-beam association. In beam selection with GoB, users select the appropriate beam from a set of pre-defined beams and the overhead during the beam selection process is a common challenge in this area. In this paper, we propose an Advantage Actor Critic (A2C) learning-based framework to improve the GoB and the beam selection process, as well as optimize transmission power in a mmWave network. The proposed beam selection technique allows performance improvement while considering transmission power improves Energy Efficiency (EE) and ensures the coverage is maintained in the network.
We further investigate how the proposed algorithm can be deployed in a Service Management and Orchestration (SMO) platform. 
Our simulations show that A2C-based joint optimization of beam selection and transmission power is more effective than using Equally Spaced Beams (ESB) and fixed power strategy, or optimization of beam selection and transmission power disjointly. 
Compared to the ESB and fixed transmission power strategy, the proposed approach achieves more than twice the average EE in the scenarios under test and is closer to the maximum theoretical EE.\looseness=-1 
\end{abstract}



\section{Introduction}
\label{intro}

mmWave technology significantly empowered the fifth-generation wireless networks (5G) due to the availability of ultra-wide frequency bands in the spectrum bands above 6 GHz. mmWave is also expected to take an important role in Beyond 5G (B5G) technologies, and even the range of THz communications is considered as a potential enabler in the next generations \cite{thz}. The gains of using mmWave can only be unlocked with the use of large antenna arrays that generate ultra-narrow beams and overcome propagation losses. This gives beam management a central role in modern wireless systems.\looseness=-1 

Beam selection is the process to establish the best beam pair for the transmitter and the receiver among the available beams on both sides. The exhaustive search over all possible beam pairs could not be efficient in some cases, as in high mobility scenarios. The overhead in this process could consume a significant fraction of the  channel coherence interval and leave little time for utilization \cite{beam_selection01}. Different solutions were applied to address this problem, such as the use of out-of-band information \cite{beam_selection01}, hierarchical beamforming \cite{hierarchical_bf}, and data-driven methods based on contextual information \cite{bf_data01}.

While efficient beam management procedures are envisioned, global goals for wireless network systems, such as Energy Efficiency (EE), remain on the horizon. EE was pointed out by Next Generation Mobile Networks as a key factor in minimizing the Total Cost of Ownership (TCO) and the environmental footprint of the network infrastructure \cite{ngmn_5g}. The commitment to EE improvement goals tends to stay on focus for B5G. In addition, Open Radio Access Network (O-RAN) signatory operators have recently released their technical priorities about EE \cite{oran_ee}. Not only efficient hardware and software designs are targeted, but there is a focus on EE performance indicators to be reported at different system levels, and on taking advantage of Machine Learning (ML) algorithms to automate its improvement.

Intelligent resource management through ML algorithms is seen as a way to achieve the needs of future wireless networks \cite{ai_wireless}. Automated network optimization gained priority in recent virtualized and disaggregated RAN architectures, such as Cloud RAN and Open RAN. The abstraction layers designed for network control, such as the Service Management and Orchestration (SMO) platform, are responsible for that. SMO allows the use of RAN control applications (rApps) and could be implemented on top of Open RAN, Cloud RAN or purpose-built RAN environments, where it could assume Self-Organizing Network (SON) functions\cite{ericsson_som}. Furthermore, O-RAN Alliance sees the Grid-of-Beams (GoB) optimization through ML models as one of the main use cases for MIMO and beamforming-related applications to be deployed on the SMO \cite{oran_mimo}. It could be used to implement intelligent control over beamforming functions, allowing a flexible and dynamic optimization of those parameters to address the beam selection, for instance.\looseness=-1 

This paper builds upon the idea of local feature optimization to contribute to a global goal. Based on the GoB optimization use case, a Reinforcement Learning (RL) solution is proposed in a way to select the best beam subsets and reduce the overhead of the beam selection process. The Advantage Actor Critic (A2C) framework is applied to perform the joint optimization of beam selection and gNB transmission power, targeting to improve EE without producing coverage roles. In particular, intrinsic information from the network (UE SINR levels) is leveraged to feed the RL algorithm. The algorithm will act over the GoB and power configurations according to an objective function designed to reward EE and penalize coverage holes. In this scenario, the use of an ML model can provide flexible design, improved performance, and feasible implementation in the SMO.


The main contribution of this work is the proposed coverage-aware A2C-based beam selection method for energy-efficient mmWave networks. In addition, we investigate how the proposed algorithm can be deployed in the SMO to interact with recent RAN architectures.

The remaining sections are organized as follows. In Section~\ref{related_work}, the related work in the area is presented. Section~\ref{system_model} details the system model. Section~\ref{proposed_approach} describes the proposed EE and coverage-aware beam selection algorithm. The details about the simulations and results are shown in Section~\ref{results}. Finally, Section~\ref{conclusions} concludes this work.


\section{Related Work}
\label{related_work}

ML models have been applied to solve a wide range of wireless network problems including beam management procedures \cite{bf_data01}, \cite{ai_wireless}, \cite{yujie_ra_bm} - \cite{drl_power_beam}.
In order to leverage the gains of data-driven methods, many solutions proposed the use of external information sources to feed ML algorithms and take better beam management decisions. In \cite{bf_data01}, the authors demonstrated the advantages of using GPS information to support the beam selection process in a vehicle-to-infrastructure (V2I) scenario.\looseness=-1  

The inherent error associated with geolocation data was targeted in \cite{yujie_ra_bm}, which explored the Uncertainty K-Means (UK-means) algorithm to improve the accuracy in user grouping during the beam association. In \cite{yujie_beam_management}, the joint sensing and communication paradigm was extended into a vision-aided approach. Satellite images, beyond location information, were used to support beam management decisions. Moreover, a federated learning model was proposed to address the beam selection problem in \cite{bf_fl}. The authors have showed that the collaborative learning process improved performance on throughput and best beam pairs detection compared to their baselines.\looseness=-1 



Beamforming codebook design using clustering algorithms and RL methods was the focus of \cite{drl_codebook}. The efficient codebook design was proved to provide good performance with a reduced number of beams, thus addressing the overhead in the beam selection problem. In \cite{drl_power_beam}, Deep Q-Networks (DQN) were applied to jointly solve beamforming, power control, and interference coordination problem, but no EE analysis was conducted. \looseness=-1 

Different than existing works, we propose a coverage-aware intelligent beam selection method to improve EE in mmWave networks. The proposed method investigates the trade-off between EE metrics and mmWave coverage. Moreover, this work discusses how the proposed intelligent beam selection method can be implemented in an SMO platform.


\section{System Model}
\label{system_model}

\subsection{Network Model}

The network configuration used in this work consists of a 3D beamforming capable gNB and ${N_{UE}}$ stationary User Equipment (UE). We consider analog beamforming and assume a UE could experience line of sight (LOS) or non-line of sight (NLOS) propagation conditions. Finally, the presence of UDP traffic in the downlink direction is considered. 

\subsection{Grid-of-Beams Model}

The 3D beamforming capability is represented by a regular GoB, that can be described through the set of its elevation angles $\Theta = \{\theta_1, \theta_2, …, \theta_{N_{\Theta}}\}$ and azimuth angles $\Phi = \{\phi_1, \phi_2, …, \phi_{N_{\Phi}}\}$. $\theta$ and $\phi$ are, respectively, the elevation and azimuth angles of the maximum gain in a beam. Each beam is then described by the pair $(\theta_i, \phi_j)$, where $i \in [1, N_{\Theta}]$ and $j \in [1, N_{\Phi}]$. $N_{\Theta}$ and $N_{\Phi}$ are the number of values that $\theta$ and $\phi$ could assume. The maximum number of beams available to the system, here defined as $N_{B}$, is $N_{B} = N_{\Theta} \cdot N_{\Phi}$. The full GoB, named $\Upsilon$, is obtained in terms of $\Theta$ and $\Phi$: if $\Upsilon = \Theta \times \Phi$, then:\looseness=-1 

\begin{equation}
\Upsilon = 
\begin{bmatrix}
    (\theta_1, \phi_1)        & (\theta_1, \phi_2)        & \dots  & (\theta_1, \phi_{N_{\Phi}}) \\
    (\theta_2, \phi_1)        & (\theta_2, \phi_2)        & \dots  & (\theta_2, \phi_{N_{\Phi}}) \\
    \vdots                    &       \vdots              & \ddots &        \vdots             \\
    (\theta_{N_{\Theta}}, \phi_1) & (\theta_{N_{\Theta}}, \phi_2) & \dots  & (\theta_{N_{\Theta}}, \phi_{N_{\Phi}})
\end{bmatrix}
\end{equation}

\begin{figure}[htbp]
\vspace{-0.3cm}
\centerline{\includegraphics[scale=0.2]{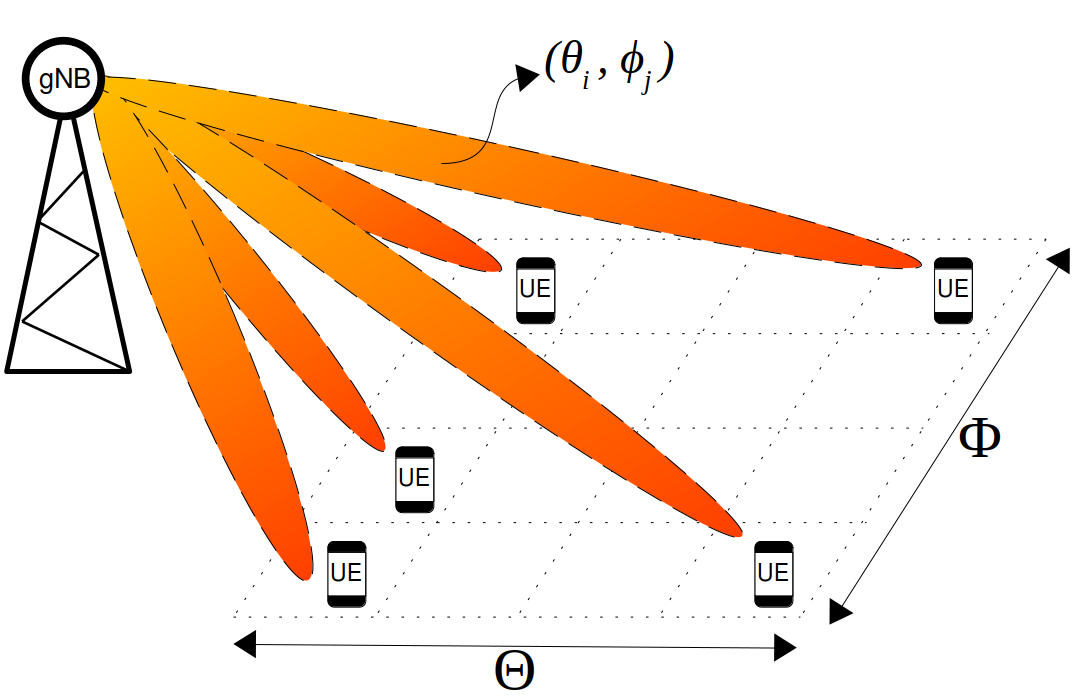}}
\vspace{-0.3cm}
\caption{Conceptual design of the 3D GoB beamforming model.}
\label{beam_grid}
\end{figure}

The 3D GoB beamforming model is depicted in Fig.~\ref{beam_grid}. The overhead in the beam selection process derives from the requirement to exchange pilot signals and evaluate the quality of each beam pair. In highly changing scenarios, it could impact the initial access and mobility procedures and consume channel utilization time. Due to this overhead, constraints on the use of $N_{B}$ beams in the beam selection could be imposed. A subset of the GoB composed of $K_{B}$ beams, where $K_{B} < N_{B}$, must be required in some cases. 
This constraint can be attended if we limit the number of elevation and azimuth angle options to a subset of $\Theta$ and $\Phi$: $\Theta'$ and $\Phi'$, where $\Theta' \subset \Theta$ and $\Phi' \subset \Phi$. Consider that $K_{B} = K_{\Theta} \cdot K_{\Phi}$, where $K_{\Theta}$ and $K_{\Phi}$ are, respectively, the number of options that each beam elevation angle $\theta$ and azimuth angle $\phi$ could assume after the constraint, so as $K_{\Theta} = |\Theta'|$ and $K_{\Phi} = |\Phi'|$.

\section{Proposed Approach}
\label{proposed_approach}

This work proposes joint optimization of beam selection and gNB transmission power, driven by EE and coverage awareness. The aim is to use GoB considering overhead constraints in the beam selection process in mmWave networks.


For the case of $K_{\Theta} < N_{\Theta}$ and $K_{\Phi} < N_{\Phi}$, $\Theta'$ and $\Phi'$ should be determined. One solution is to sample $\Theta$ and $\Phi$ according to $K_{\Theta}$ and $K_{\Phi}$, and consider equally spaced values for $\theta_i$ and $\phi_j$ within its ranges. This approach,  referred to as Equally Spaced Beams (ESB), chooses $\Theta'$ and $\Phi'$ to provide fair spacial coverage for the network area.

\begin{figure}[htbp]
\centerline{\includegraphics[scale=0.22]{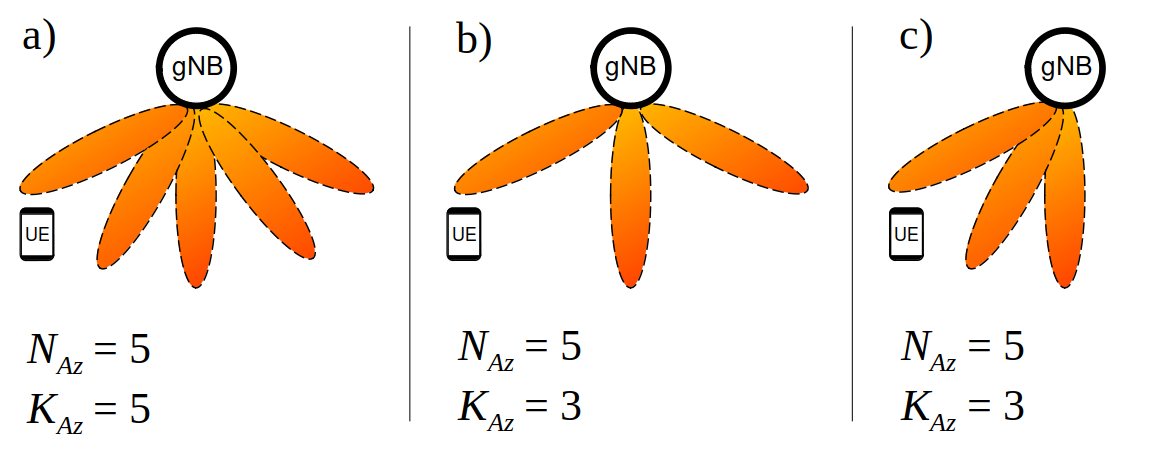}}
\vspace{-0.3cm}
\caption{Conceptual azimuth view of GoB subsets. a) $K_{\Phi} = N_{\Phi}$, then $\Phi' = \Phi$: the full GoB is in use; b) $K_{\Phi} < N_{\Phi}$: $\Phi'$ is chosen from Equally Spaced Beams (ESB) to provide fair coverage; c) $K_{\Phi} < N_{\Phi}$: $\Phi'$ is chosen from the beams that could provide the best service for the UE in this scenario. }
\label{beam_subset}
\end{figure}

Another option is to determine $\Theta'$ and $\Phi'$ according to the characteristics of the scenario.
Moreover, a dynamic choice over the subset of $\Theta$ and $\Phi$ could prove to be more efficient. 
Fig.~\ref{beam_subset} illustrates different ways to determine GoB subsets in the case of overhead constraints. 

\vspace{-0.3cm}
\subsection{The Advantage Actor Critic (A2C) Framework}

The optimization of beam selection and transmission power is done using the RL framework A2C. The foundational goal in RL is to learn a policy based on the sequential actions and environment states that maximize a cumulative future reward. If an action has a high expected reward, it should be given a higher probability of selection for an observed state. A2C differs from other Deep Reinforcement Learning (DRL) frameworks because it proposes a clear separation between two tasks performed by the agent. Each block within the agent (actor and critic) has its dedicated neural network engine to perform the corresponding task.
The critic's task is to judge the actor's decisions based on the current state of the environment and the outcomes provided as a reward. The actor's task is to define the best action to be performed on the environment based on the current state and the policy values sent by the critic \cite{a2c}.\looseness=-1 

\subsection{Markov Decision Process (MDP) Formulation}

\subsubsection{State Space}

The proposed environment is the network model described in Section~\ref{system_model} and presented to the algorithm through the state space. $s_t$ consists of the SINR values $\Gamma$ observed for each UE, quantized and truncated over the range $[0,10]$, here represented as $\hat{\Gamma}$. 

For each UE,
\begin{equation}
\hat{\Gamma}_i = \begin{cases} 0, $ for $ \Gamma_i <= 0  \\ \lfloor \Gamma_i \rceil, $ for $ 0 < \Gamma_i < 10 \\ 10, $ for $ \Gamma_i >= 10\end{cases}
\end{equation}

Thus, the state space is formulated as:
\begin{equation}
s_t = [\hat{\Gamma}_1, \hat{\Gamma}_2, ..., \hat{\Gamma}_{N_{UE}}].
\end{equation}

\subsubsection{Action Space}

Following the GoB model described in Section~\ref{system_model}, the action space is here defined as:

\begin{equation}
a_t = [\Theta_l', \Phi_m', \delta_n],
\end{equation}

\noindent where $\Theta_l' \subset \Theta$ and $\Phi_m' \subset \Phi$.  $l$ and $m$ are subset indexes so as $l \in [1, L]$ and $m \in [1, M]$, where $L$ and $M$ represent the number of different subset options from $\Theta$ and $\Phi$ that account for $K_{\Theta}$ and $K_{\Phi}$ constraints. Here, $L$ and $M$ are defined as: $L = \binom{N_{\Theta}}{K_{\Theta}}$ and $M = \binom{N_{\Phi}}{K_{\Phi}}$. Consider that $\Theta^L = \binom{\Theta}{K_{\Theta}}$ and $\Phi^M = \binom{\Phi}{K_{\Phi}}$, so $\Theta^L$ and $\Phi^M$ could represent the sets of subsets from $\Theta$ and $\Phi$ that account for $K_{\Theta}$ and $K_{\Phi}$ constraints. Hence, $\Theta_l' \in \Theta^L$ and $\Phi_m' \in \Phi^M$.

$\delta_n$ accounts for a power level change over the baseline power level $P_B$, chosen from the set of power level change options $\Delta$. $n$ is a change index, so as $n \in [1,N_P]$, and $N_P = |\Delta|$.\looseness=-1 

\subsubsection{Reward Function}  
The reward function is proposed in a way to reinforce learning towards the best achievable EE and to avoid the presence of coverage holes in the scenario. It is defined as follows:

\begin{equation}
r_t = {\frac{\varepsilon}{\varepsilon_{MAX}}} - \omega_C \cdot \rho,
\end{equation} 

\noindent where $\varepsilon$ represents the EE in terms of gNB throughput ($\tau$) and transmission power ($P$), so as $\varepsilon = \frac{\tau}{P}$. Here, $\varepsilon_{MAX}$ is the maximum theoretical EE, derived from the data rate provided by the server application in the system ($\tau_{MAX}$) and the minimum amount of power available to be selected ($P_{MIN}$), i.e.  $\varepsilon_{MAX} = \frac{\tau_{MAX}}{P_{MIN}}$. Furthermore, $\rho$ determines the UE outage ratio, based on an SINR threshold $\hat{\Gamma}^T$ and $\rho = \frac{n_{UE}}{N_{UE}}$, where $n_{UE}$ is the number of UE served by the gNB with $\hat{\Gamma} <= \hat{\Gamma}^T$. Since the second term in the reward function determines the penalties according to the UE outage ratio, $\omega_C$ is used as an out-of-coverage density factor weight, customizable to tune the severity of the penalty and balance the two goals within the reward function: EE and coverage awareness. In summary, $r_t$ is proportional to EE, and it seeks to maintain a greater number of UEs above the SINR threshold.   

\subsection{Baseline Algorithms} 
Two other versions of the MDP described above are evaluated as baselines. The first one, named A2C-P, intends to cover the case where only power optimization is conducted by the algorithm. In this case, ESB is used to define the constrained GoB. It uses the same state space and reward function presented above, however, the action space is defined as:\looseness=-1 

\begin{equation}
a^P_t = [\delta_n].
\end{equation}

A second variant named A2C-B is proposed with the intent to cover the case where only beam selection optimization is conducted. In this case, the baseline power level is used. As A2C-P, it uses the same state space and reward function as A2C, but the action space is defined as:

\begin{equation}
a^B_t = [\Theta_l', \Phi_m'].
\end{equation}

Finally, we consider the use of ESB and fixed transmission power as a third baseline, covering the case of no dynamic optimization.\looseness=-1 

\subsection{Algorithm Deployment}

rApps are applications designed to automate and control RAN functions and run on top of the SMO platform. rApps are enabled by the Non-Real Time RAN Intelligent Controller (Non-RT RIC), an SMO entity, and operate in control loops of 1 second or more \cite{oran_arch}. In accordance with the GoB optimization use case presented in \cite{oran_mimo}, the approach proposed in this work may be deployed as an rApp in the SMO. Note that, the implementation as an rApp is feasible because the global GoB and power optimization proposed here do not require near real-time feedback and interventions.

\begin{figure}[htbp]
\centerline{\includegraphics[scale=0.09]{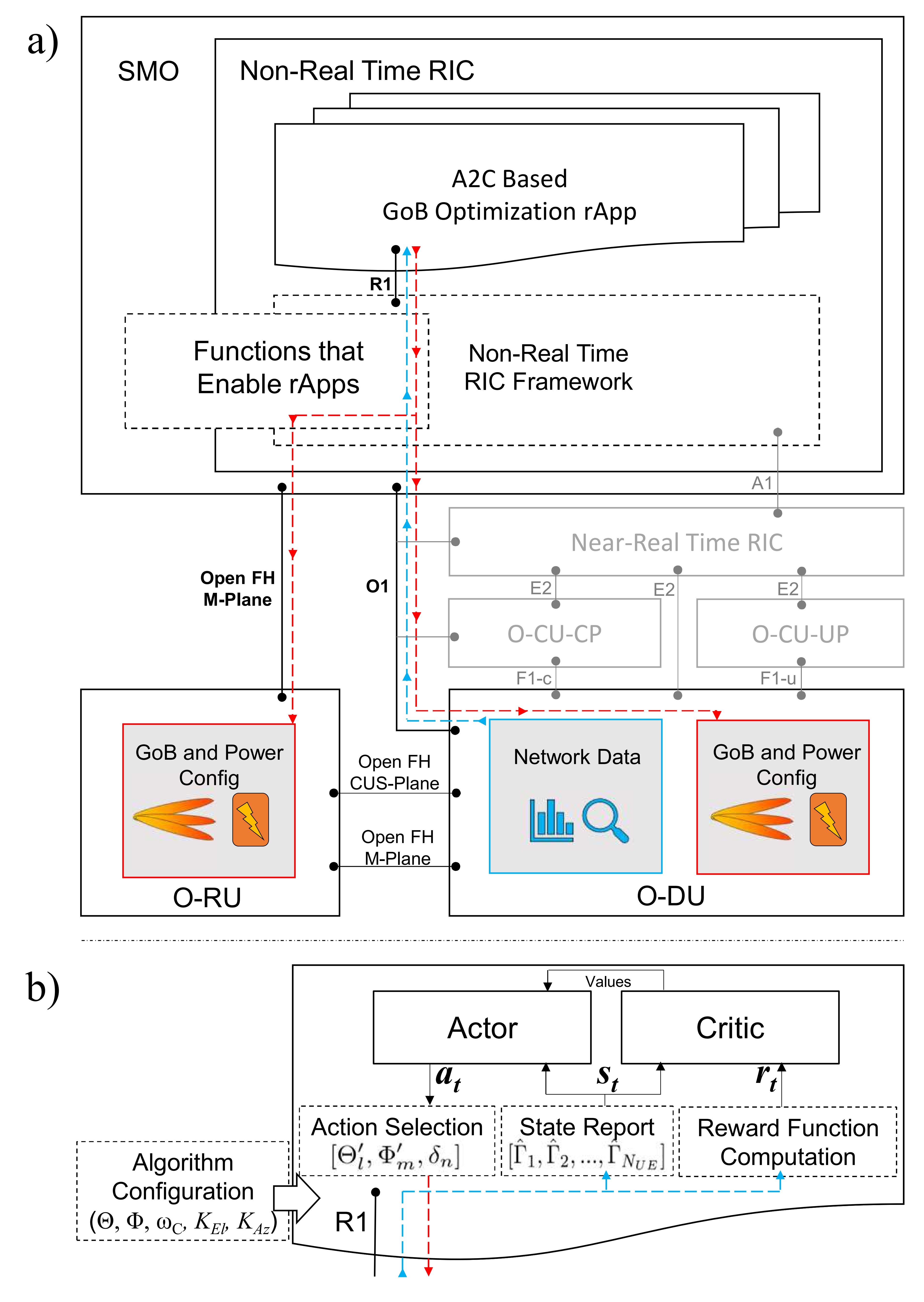}}
\vspace{-0.5cm}
\caption{The proposed solution deployed as an rApp in the SMO.
a)  Data flow through the rApp and Open RAN entities
(SMO: Service Management and Orchestration; RIC: RAN Intelligent Controller; O-RU: Open Radio Unit; O-DU: Open Distributed Unit; O-CU: Open Centralized Unit; R1, O1, A1, E2, Open FH M-Plane, and Open FH CUS-Plane are open interfaces; F1-c and F1-u are 3GPP-defined interfaces);
b) A2C based rApp internal functioning ($a_t$: action taken at time $t$; $s_t$: state reported at time $t$; $r_t$: reward value at time $t$).
}
\label{actor_critic}
\end{figure}

Fig.~\ref{actor_critic}-a) shows a potential deployment where the proposed solution is used as an rApp in the SMO for Open RAN architecture. In O-RAN, network data gathered by the Open Decentralized Unit (O-DU) is sent to the SMO and Non-RT RIC Framework through O1 interface. Non-RT RIC Framework uses R1 communication to transfer the data to the rApp. 

In our case, the A2C-based GoB and power configurations (actions) can be applied to the Open Radio Unit (O-RU) through Open FH M-Plane interface, and to O-DU through O1 interface. Note that, Non-RT RIC rApps are not limited to the Open RAN environment. With the use of proper interfaces, similar deployment could be done on top of Cloud RAN or purpose-built network architectures. 

Fig.~\ref{actor_critic}-b) details the functioning of the proposed scheme when it is developed as an rApp. The algorithm configuration is done based on $\Theta$, $\Phi$, $\omega_C$, $K_{\Theta}$ and $K_{\Phi}$. In each loop, data collected from O-DU will enable the reward function computation (feeds the critic) and the state report (feeds the actor and the critic). The action selection will be done by the actor and then sent as an rApp policy to be applied to O-RU and O-DU.\looseness=-1 

\section{Simulations and Results}
\label{results}

\subsection{Simulation Environment and Settings}
We implement our proposed scheme using the 5G LENA module for ns-3 \cite{5g_lena}. The main system parameters for the simulation environment are summarized in Table \ref{sys_param}.  

Each simulation is set to last 150 ms, where only 100 ms are used for data traffic purposes. Different scenarios were set according to the intended number of UE, since $N_{UE}=\{5, 10, 15, 20\}$. The results shown in Section \ref{results} refer to the average value and confidence interval of 40 different seed runs for each scenario configuration. In each run, new channel parameters and UE distribution are set.

A regular GoB where $N_{\Theta} = 3$ and $N_{\Phi} = 5$, so as $N_{B} = 15$, is considered here. The set of available angles for each axis are the following: $\Theta = \{-15^{\circ}, -45^{\circ}, -60^{\circ}\}$ and $\Phi = \{0^{\circ}, 45^{\circ}, 90^{\circ}, 135^{\circ}, 180^{\circ}\}$. $K_{\Theta} = 1$ and $K_{\Phi} = 3$ are defined considering a constraint of $K_{B} = 3$. For the A2C based approach, consider $L = 3$ and $M = 10$. 
For the ESB strategy with $K_B = 3$, $\Theta' = \{-45^{\circ}\}$ and $\Phi' = \{0^{\circ}, 90^{\circ}, 180^{\circ}\}$ are considered. Both ESB strategies use the baseline power level ($P_B$), so as A2C-B. 
$P_B$ is set as 35 dBm. Five power level change options $\delta$ are defined, so as $N_P = 5$. The set of power level change options is the following: $\Delta = \{-3, -1.5, 0, +1.5, +3\}$  dB.

A2C implementation was done using the python package Stable Baselines 3 \cite{stables_baselines_3}. Each A2C episode corresponds to a new simulation round, fed with the action set by A2C, and that returns the states and reward values needed to feed the RL algorithm's next iteration. The convergence curve for the reward values obtained by A2C in the case where $N_{UE} = 5$ is depicted in Fig.~\ref{convergence_curve}. 

\begin{table}[htbp]
\centering
\caption{System Parameters}
\begin{tabular}{@{}ll@{}}
\toprule
Parameter                    & Value                         \\ \midrule
Simulation Area              & 50m x 50m                     \\
gNB Antenna Array            & 4x32                          \\
Number of gNB               & 1                             \\
Number of UE                & \{5, 10, 15, 20\}             \\
Carrier Frequency            & 28GHz                         \\
Bandwidth                    & 100MHz                        \\
Numerology                   & 2                             \\
Component Carriers           & 1                             \\
Propagation Model            & 3GPP TR 38.900 Urban Macro    \\
gNB TX Power                 & \{32, 33.5, 35, 36.5, 38\}dBm \\
UE TX Power                  & 10dBm                         \\
Traffic Type                 & DL UDP                        \\
Packet Size                  & 400KB                         \\
Data Rate (User Application) & 1Mbps                         \\ \bottomrule
\end{tabular}
\label{sys_param}
\vspace{-0.5cm}
\end{table}

\begin{figure}[htbp]
\includegraphics[scale=1]{convergence_curve_5_ues.png}
\vspace{-0.5cm}
\caption{A2C convergence curve for the scenario with $\omega_C = 1$ and $N_{UE} = 5$. The shaded area represents the range of the confidence interval out of 40 simulation runs.}
\label{convergence_curve}
\vspace{-0.2cm}
\end{figure}

\subsection{Results}
\begin{figure}[htbp]
\centerline{\includegraphics{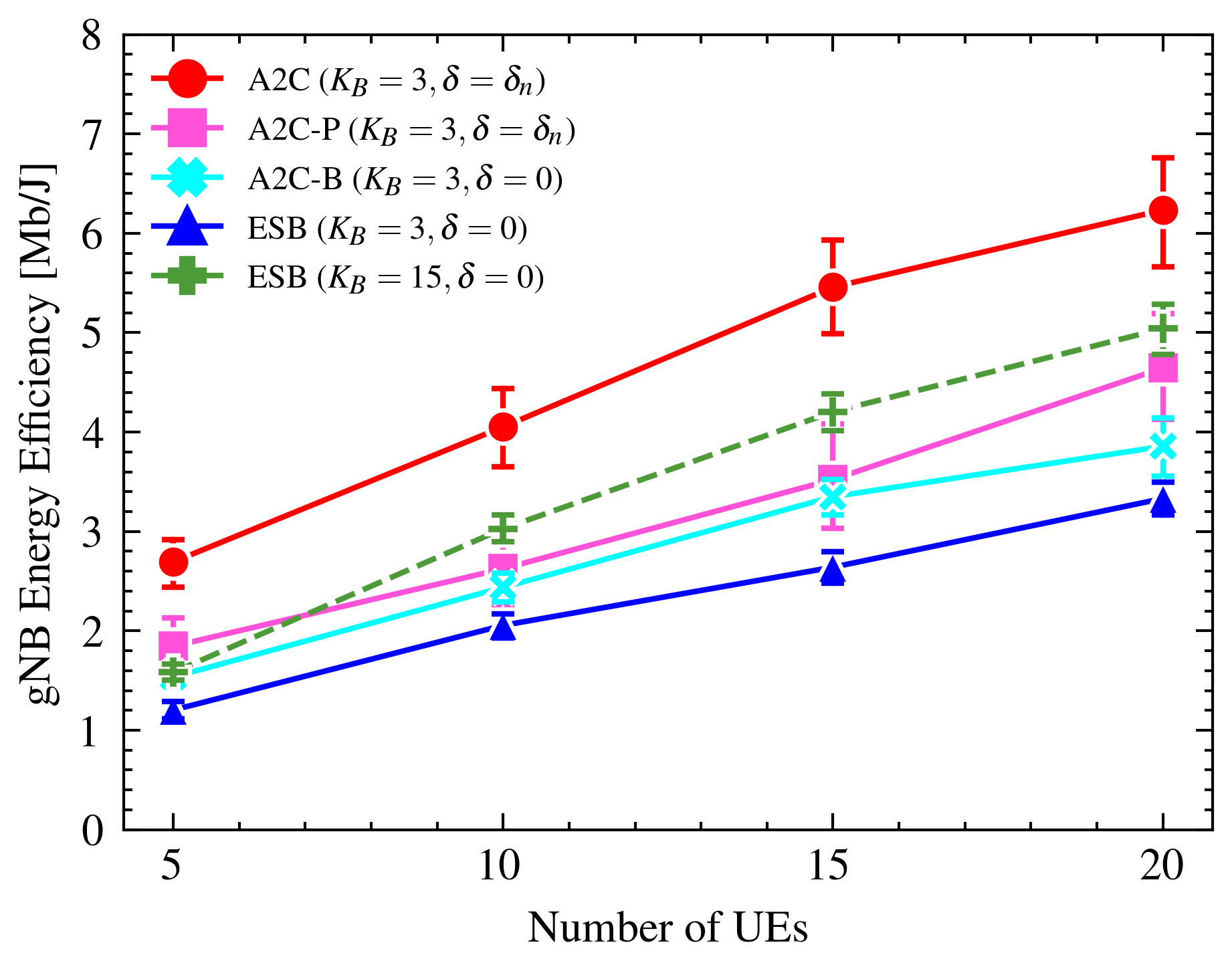}}
\vspace{-0.5cm}
\caption{gNB Energy Efficiency. $K_B$ is the number of beams in the grid. $\delta$ is the power level change from the baseline power level.} 
\label{ef}
\end{figure}

Fig.~\ref{ef} shows the EE values obtained for the proposed approach (A2C) when compared to the ESB for $K_B = 3$ and $K_B = 15$. The cases with only the power or the beam subset optimization (A2C-P and A2C-B) are also shown. $K_B = 15$ represents the case where $\Upsilon' = \Upsilon$, and the beam search over the full GoB is performed.   

A2C outperforms ESB for the same number of beams. More importantly, it can improve EE even when compared to the case of greater $K_B$ with no power optimization. The ratio of $\varepsilon_{MAX}$ achieved by each strategy is summarized in Table~\ref{ef_gains_tab}.

\begin{table}[htbp]
\vspace{-0.2cm}
\centering
\caption{Max Energy Efficiency Ratio \\
{$\varepsilon / \varepsilon_{MAX}$} }
\begin{tabular}{@{}lccccc@{}}
\toprule
                   &  \multicolumn{4}{c}{$N_{UE}$}                &        \\
Algorithm          & $5$      & $10$      & $15$      & $20$      &  Avg.  \\
\midrule
\textbf{A2C ($K_B=3, \delta=\delta_n$)}       & 0.74     & 0.55      & 0.50      & 0.43      &  \textbf{0.56}  \\
A2C-P ($K_B=3, \delta=\delta_n$)   & 0.51     & 0.36      & 0.32      & 0.32      &  0.38  \\
A2C-B ($K_B=3, \delta=0$)   & 0.42     & 0.34      & 0.31      & 0.27      &  0.33  \\
\textbf{ESB ($K_B=3, \delta=0$)}     & 0.33     & 0.28      & 0.24      & 0.23      & \textbf{0.27}  \\
ESB ($K_B=15, \delta=0$)    & 0.44     & 0.42      & 0.39      & 0.35      &  0.40  \\
\bottomrule
\vspace{-0.6cm}
\end{tabular}
\label{ef_gains_tab}
\end{table}

\begin{figure}[tbp]
\vspace{-0.3cm}
\centerline{\includegraphics{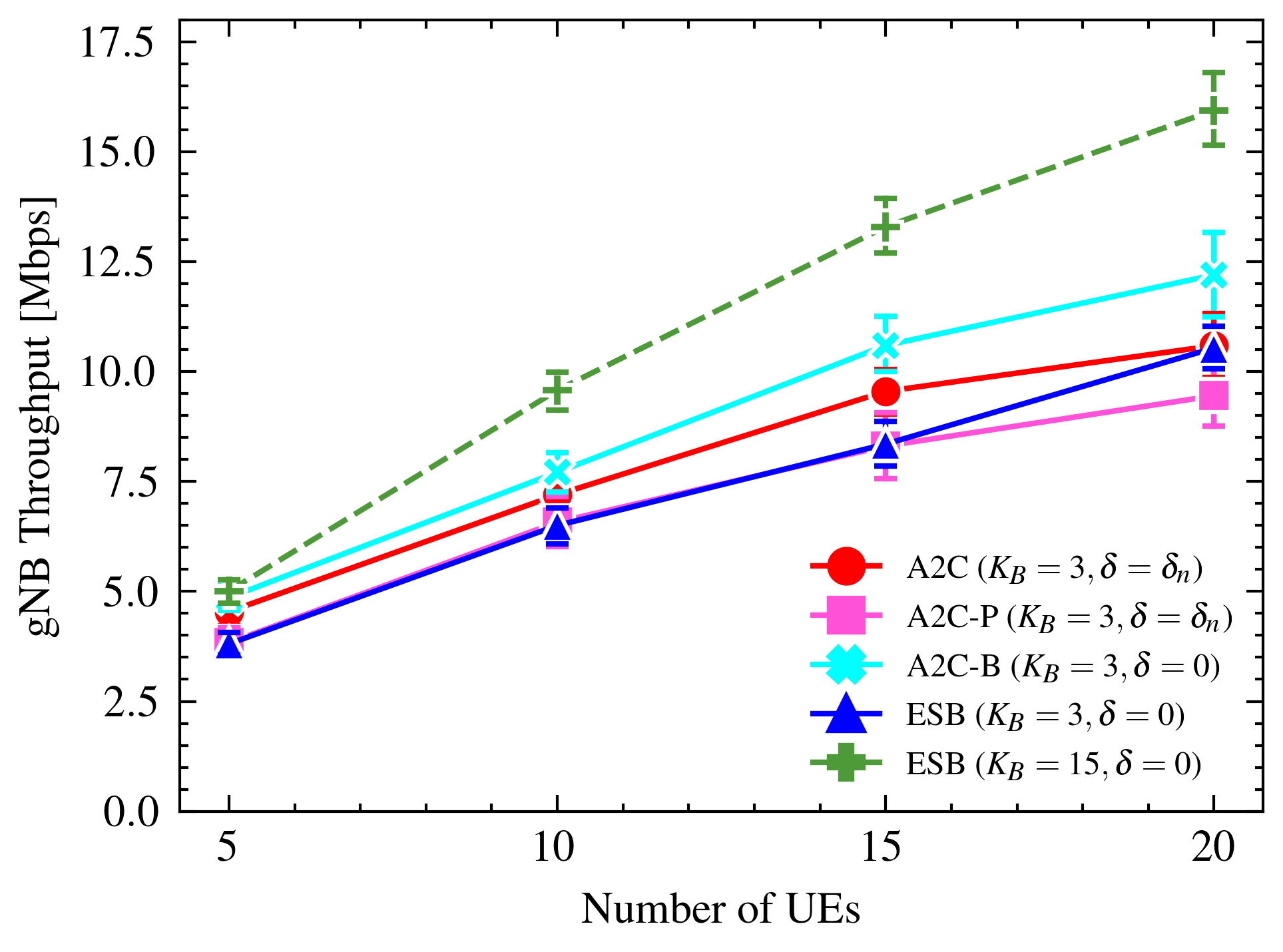}}
\vspace{-0.5cm}
\caption{gNB Throughput. $K_B$ is the number of beams in the grid. $\delta$ is the power level change from the baseline power level.}
\label{gnb_tp}
\end{figure}

For the same number of beams ($K_B = 3$), the use of A2C does not cause a significant impact on throughput, as shown in  Fig.~\ref{gnb_tp}. 
The results for the complete original GoB ($K_B = 15$) are shown to demonstrate the achievable performance when no constraints to $N_B$ are applied. In the case of throughput, $K_B = 15$ achieves higher values due to the gains in SINR derived from the use of more beams. However, it is not compliant with the constraints defined in our problem ($K_B = 3$). $K_B = 15$ with fixed power is also less energy-efficient than the proposed solution.

\begin{figure}[tbp]
\centerline{\includegraphics{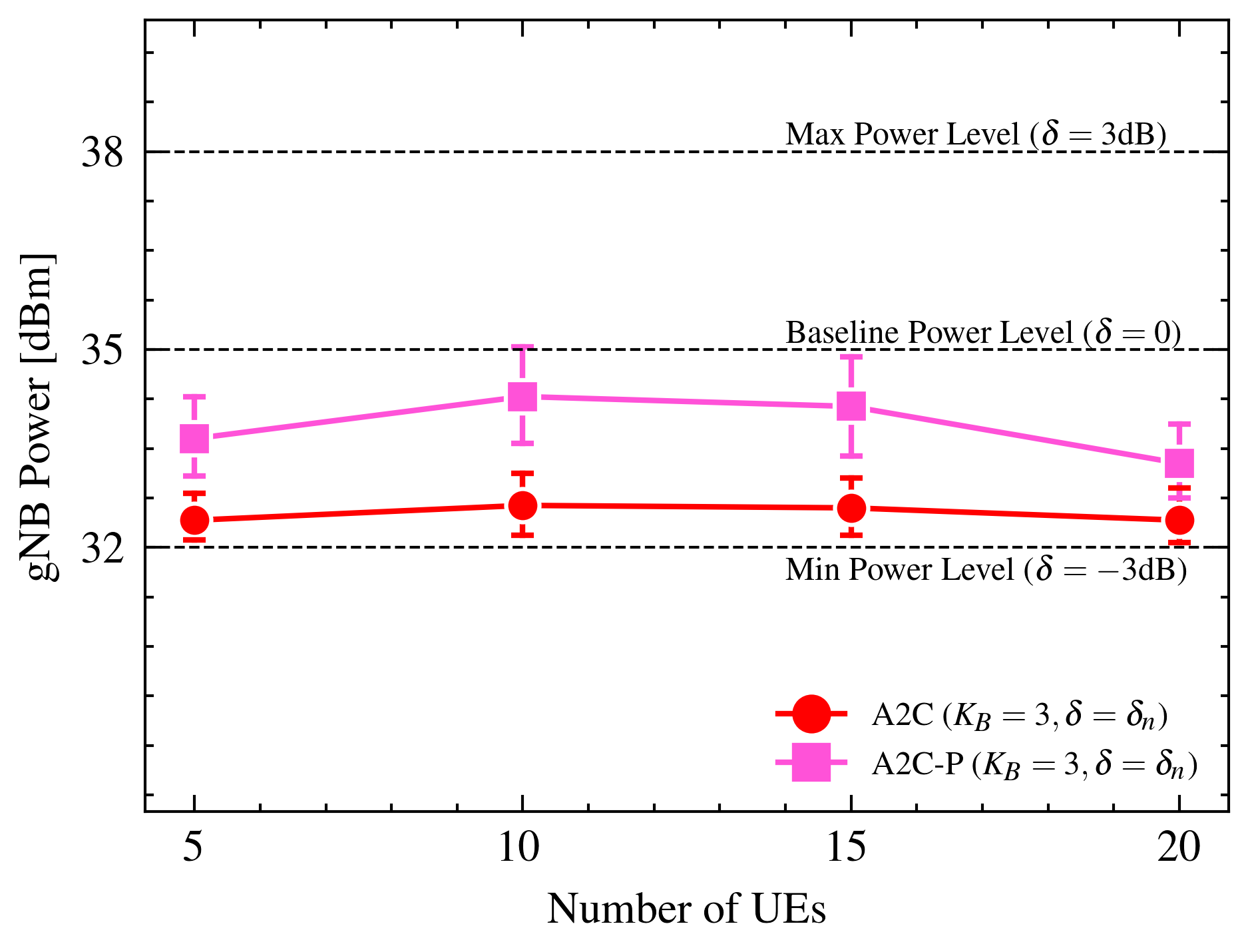}}
\vspace{-0.5cm}
\caption{gNB Power Usage. $K_B$ is the number of beams in the grid. $\delta$ is the power level change from the baseline power level.}
\label{power_dbm}
\vspace{-0.5cm}
\end{figure}

Fig.~\ref{power_dbm} shows that A2C could converge to minimize transmission power at the gNB. If only power optimization is performed (A2C-P), throughput is not improved, and the power levels in use are diminished but are higher than for the global A2C case (Fig.~\ref{power_dbm}). Thus, A2C-P presents poor EE improvement. When only the beam subset optimization is performed (A2C-B), throughput is improved, but since the power is fixed and not optimized, EE gains are discrete. Moreover, A2C-P and A2C-B have similar performances on EE. It is the joint optimization of beam selection and transmission power that produces the best EE results. The proposed approach leveraged the gains from the beam subset optimization with throughput to minimize transmission power and improve EE at the gNB.\looseness=-1 

\begin{figure}[htbp]
\vspace{-0.3cm}
\centerline{\includegraphics{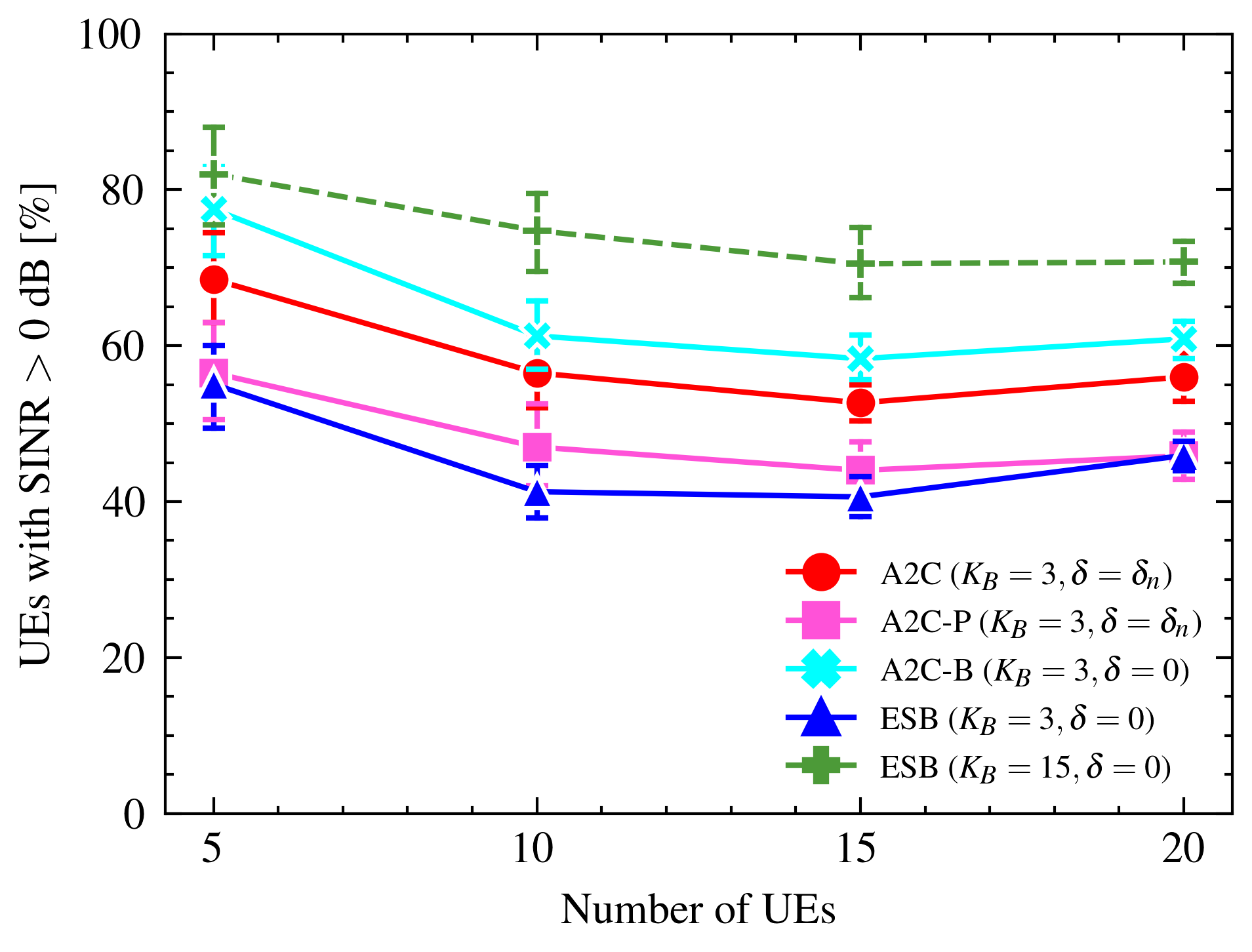}}
\vspace{-0.5cm}
\caption{Percentage of UE with SINR Above 0 dB. $K_B$ is the number of beams in the grid. $\delta$ is the power level change from the baseline power level.}
\label{above0sinr}
\end{figure}

As shown in Section~\ref{proposed_approach}, the reward function of the proposed approach is designed to improve EE and avoid the presence of coverage holes. Fig.~\ref{above0sinr} shows how the proposed approach reduces the number of UE with $\hat{\Gamma} < 0$ in the scenario when compared to the ESB case with the same number of beams. It happened even though the use of A2C translated into lower levels of power applied in the system.

\vspace{-0.2cm}




\section{Conclusions}
\label{conclusions}

This paper showed the design and evaluation of an A2C-based strategy for GoB optimization. The A2C framework was applied to promote EE improvement and coverage awareness, deciding jointly on the beam selection and transmission power at the gNB. A deployment option on an SMO platform was also shown.
The proposed approach is able to optimize GoB beamforming and introduce significant gains in EE. The algorithm could place good decisions in terms of beam subset choice and transmission power with the use of intrinsic information only (UE SINR levels). The gains from the beam selection were leveraged to minimize the transmission power in the system. No throughput or coverage losses were required to improve EE, and the balance between EE and coverage was explored within the reward function, which accounts for the flexibility of this proposal. Future directions for this work include the use of the proposed approach on multi-gNB and mobility scenarios.


\section*{Acknowledgment}
This work is partly supported by MITACS and NSERC programs. The work of Ycaro Dantas is partly supported by the Vector Scholarship in Artificial Intelligence, provided through the Vector Institute.



\begin{thebibliography}{00}

\bibitem{thz} C. Chaccour, M. N. Soorki, W. Saad, M. Bennis, P. Popovski and M. Debbah, ``Seven defining features of terahertz (THz) wireless systems: a fellowship of communication and sensing," \textit{IEEE Communications Surveys and Tutorials}, vol. 24, no. 2, pp. 967-993, second quarter 2022.

\bibitem{beam_selection01} A. Ali, N. González-Prelcic and R. W. Heath, ``Millimeter Wave Beam-Selection Using Out-of-Band Spatial Information," \textit{IEEE Transactions on Wireless Communications}, vol. 17, no. 2, pp. 1038-1052, Feb. 2018.

\bibitem{hierarchical_bf}J. Wang, ``Beam codebook based beamforming protocol for multi-Gbps millimeter-wave WPAN systems", IEEE J. Sel. Areas Commun., vol. 27, no. 8, pp. 1390-1399, Oct. 2009.

\bibitem{bf_data01} A. Klautau, P. Batista, N. González-Prelcic, Y. Wang and R. W. Heath, ``5G MIMO Data for Machine Learning: Application to Beam-Selection Using Deep Learning," in \textit{2018 Information Theory and Applications Workshop (ITA)}, pp. 1-9, Feb. 2018.


\bibitem{ngmn_5g} NGMN Alliance, ``NGMN 5G white paper," March 2015.

\bibitem{oran_ee} Deutsche Telekom, Orange, Telefonica, TIM, and Vodafone ``Open RAN Technical Priorities, Focus on Energy Efficiency," June 2021.

\bibitem{ai_wireless} M. Elsayed and M. Erol-Kantarci, "AI-Enabled Future Wireless Networks: Challenges, Opportunities, and Open Issues," in \textit{IEEE Vehicular Technology Magazine}, vol. 14, no. 3, pp. 70-77, Sept. 2019.

\bibitem{ericsson_som} Ericsson, ``Intelligent Optimization Guide," June 2022. Available at https://www.ericsson.com/en/ran/intelligent-ran-automation.


\bibitem{oran_mimo} O-RAN Alliance Working Group 1, ``O-RAN Massive MIMO Use Cases Technical Report 1.0," July 2022. Available at https://www.o-ran.org/.


\bibitem{yujie_ra_bm} Y. Yao, H. Zhou and M. Erol-Kantarci, ``Deep reinforcement learning-based radio resource allocation and beam management under location uncertainty in 5G mmWave networks," arXiv:2204.10984, Apr. 2022.

\bibitem{yujie_beam_management} Y. Yao, H. Zhou and M. Erol-Kantarci, ``Joint Sensing and Communications for Deep Reinforcement Learning-based Beam Management in 6G," arXiv:2208.01880, Aug. 2022.


\bibitem{bf_fl} M. B. Mashhadi, M. Jankowski, T. -Y. Tung, S. Kobus and D. Gündüz, "Federated mmWave Beam Selection Utilizing LIDAR Data," in \textit{IEEE Wireless Communications Letters}, vol. 10, no. 10, pp. 2269-2273, Oct. 2021.



\bibitem{drl_codebook} Y. Zhang, M. Alrabeiah and A. Alkhateeb, ``Reinforcement learning of beam codebooks in millimeter wave and terahertz MIMO systems," \textit{IEEE Transactions on Communications}, vol. 70, no. 2, pp. 904-919, Feb. 2022.

\bibitem{drl_power_beam} F. B. Mismar, B. L. Evans and A. Alkhateeb, ``Deep reinforcement learning for 5G networks: joint beamforming, power control, and interference coordination," \textit{IEEE Transactions on Communications}, vol. 68, no. 3, pp. 1581-1592, March 2020.

\bibitem{oran_arch} M. Polese, L. Bonati, S. D'Oro, S. Basagni, and T. Melodia, ``Understanding O-RAN: architecture, interfaces, algorithms, security, and research challenges,"	arXiv:2202.01032, Feb. 2022.

\bibitem{5g_lena} Natale Patriciello, Sandra Lagen, Biljana Bojovic, Lorenza Giupponi, ``An E2E simulator for 5G NR networks", \textit{Simulation Modelling Practice and Theory (SIMPAT)}, Vol. 96, 101933, Nov. 2019.

\bibitem{a2c} M. Sewak, ``Actor-critic models and the A3C,” \textit{Deep Reinforcement Learning}, pp. 141–152, Springer, 2019.

\bibitem{3gpp_uma} 3rd Generation Partnership Project (3GPP), ``Study on channel model for frequency spectrum above 6 GHz,” 3GPP TR 38.900 version 14.2.0 Release 14, 2017.

\bibitem{stables_baselines_3} A. Raffin, A. Hill, A. Gleave, A. Kanervisto, M. Ernestus, and N. Dormann, ``Stable-Baselines3: reliable reinforcement learning implementations", \textit{Journal of Machine Learning Research}, 2021.

\end{thebibliography}
\end{document}